\documentclass{aastex}

\usepackage{epsfig}
\usepackage{lscape}
\usepackage{apjfonts}
\usepackage{mathptmx} 

\newcommand{\myemail}{s.dzib@crya.unam.mx}

\newcommand{\Msun}{$M_\odot$}

\shorttitle{The distance to HW 9 in Cepheus A}
\shortauthors{Dzib et al.}

\begin{document}

\title{VLBA determination of the distance to nearby star-forming regions\\
VI. The distance to the young stellar object HW 9 in Cepheus A}

\author{Sergio Dzib, Laurent Loinard, Luis F.\ Rodr\'{\i}guez}

\affil{Centro de Radioastronom\'{\i}a y Astrof\'{\i}sica, Universidad
Nacional Aut\'onoma de M\'exico\\ Apartado Postal 3-72, 58090,
Morelia, Michoac\'an, M\'exico (\myemail)}

\author{Amy J.\ Mioduszewski}

\affil{National Radio Astronomy Observatory, Domenici Science Operations Center,\\
1003 Lopezville Road, Socorro, NM 87801, USA}

\and

\author{Rosa M.\ Torres}

\affil{Argelander-Institut f\"ur Astronomie,  Universit\"at Bonn,\\
 Auf dem H\"ugel 71, 53121 Bonn, Germany}

\begin{abstract}
Using the Very Long Baseline Array (VLBA), we have observed the radio continuum
emission from the young stellar object HW 9 in the Cepheus A star-forming 
region at ten epochs between 2007 February and 2009 November. Due to its 
strong radio variability, the source was detected at only four of 
the ten epochs. From these observations, the trigonometric parallax of HW 9 
was determined to be $\pi$ = 1.43 $\pm$ 0.07 mas, in excellent agreement 
with a recent independent VLBA determination of the 
trigonometric parallax of a methanol maser associated with the nearby young 
stellar source HW 2 ($\pi$ = 1.43 $\pm$ 0.08 mas). This concordance in results, 
obtained in one case from continuum and in the other from line observations,
confirms the reliability of Very Long Baseline Array trigonometric parallax measurements. 
By combining the two results, we constrain the distance to Cepheus A to be 
700$_{-28}^{+31}$ pc, an uncertainty of 3.5\%. 
\end{abstract}


\section{Introduction}

The Cepheus A region is an active site of Galactic star-formation, which contains 
one of the very few well-documented examples (HW 2) of a high-mass protostellar 
system where a disk-like flattened structure has been detected (Patel et al.\ 2005). 
The distance to Cepheus A has traditionally been very uncertain with estimates ranging
from 300 pc (Migenes et al.\ 1992) to 900 pc (Moreno-Corral et al.\ 1993). Recently,
Moscadelli et al.\ (2009) obtained a direct parallax measurement based on multi-epoch
observations of a methanol maser feature associated with HW 2, using the VLBA telescope. 
The corresponding result ($d$ = 700 $\pm$ 40 pc) significantly reduced the uncertainty on the distance to 
Cepheus A, but was based on a single set of observations of a single source. In the 
present paper, the sixth of our series dedicated to Very Long Baseline Array determinations 
of distances to nearby star-forming regions (see Loinard et al. 2005, 2007, 2008; Torres 
et al.\ 2007, 2009; Dzib et al.\ 2010 for the previous papers of the series), we will
provide an entirely independent measurement of the distance to Cepheus A based on 
multi-epoch observations of  the continuum emission associated with the radio source 
HW 9.

\begin{figure}[!ht]
\begin{center}
\includegraphics[width=0.8\linewidth,angle=-90]{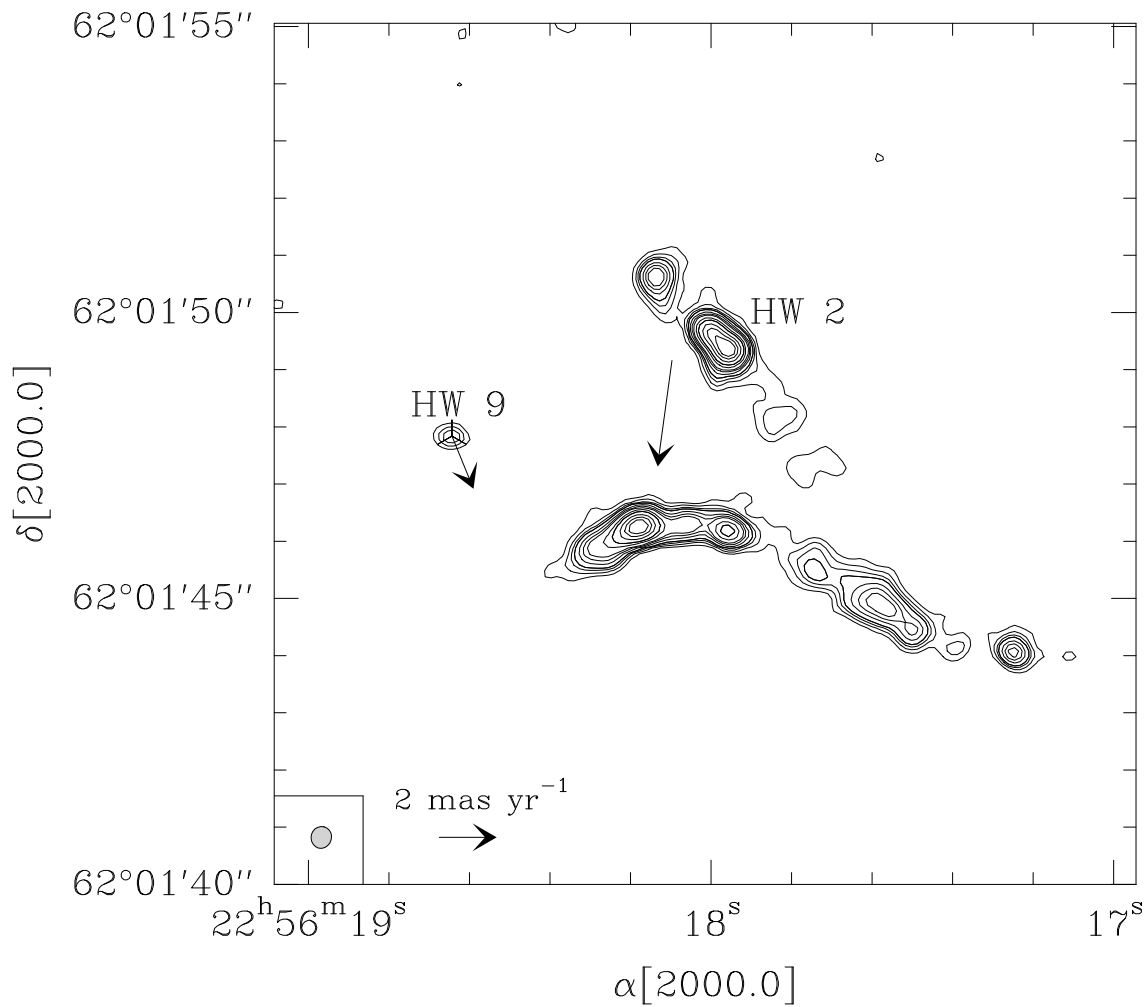}
\end{center}
\caption{Central region of Cepheus A observed with the VLA in A configuration at $\nu$ = 4.86 GHz
on 2006 February 11 (project AC810). The data were downloaded from the VLA archive, and calibrated
following standard procedures. The angular resolution is about 0.4 arcseconds, and the contours are
at -3, 3, 6, 9, 12, 18, 24, 30, 45, 60, 90, 120 and 150 times the noise level in the image ($\sigma$ = 20 
$\mu$Jy beam$^{-1}$). The arrows represent the proper motions of HW 9 and of the maser associated 
with HW 2 reported by Moscadelli et al.\ (2009).}
\label{fig:pm}
\end{figure}

HW 9 was first reported by Hughes (1991) as the ninth radio source in Cepheus A (following 
previous detections in the same region reported in Hughes \& Wouterloot\ 1984). Hughes\ (1991) 
found that the source was very compact as well as highly variable, and interpreted its radio 
emission in terms of gyrosynchrotron radiation (see also Hughes et al.\ 1995 and Garay 
et al.\ 1996). In high angular resolution maps (Figure \ref{fig:pm}), HW 9 appears as an 
isolated, featureless source located about 5$''$ to the south-east of the well-studied 
high-mass object HW 2. The visual extinction toward HW 9 is at least 23 magnitudes 
(Pravdo et al.\  2009) and could be as high as 100 magnitudes (Hughes\ 1991; Pravdo 
et al.\ 2009). As a consequence, no infrared or visual counterpart has ever been reported 
for HW 9, no spectral classification is available, and the very nature of the source remains 
very unclear. Because of its possible association with an H {\small II}  region, Hughes (1991) 
proposed that HW 9 might be a B3 star. Garay et al.\ (1996), however, argued that HW 9 is 
more likely to be a low mass stellar object because of its radio flaring activity. The observations 
reported in this paper will provide some additional constraints on the nature of this 
enigmatic source.

\section{Observations and Data Reductions}

\begin{figure}[!ht]
\begin{center}
\includegraphics[width=0.85\linewidth,angle=-90]{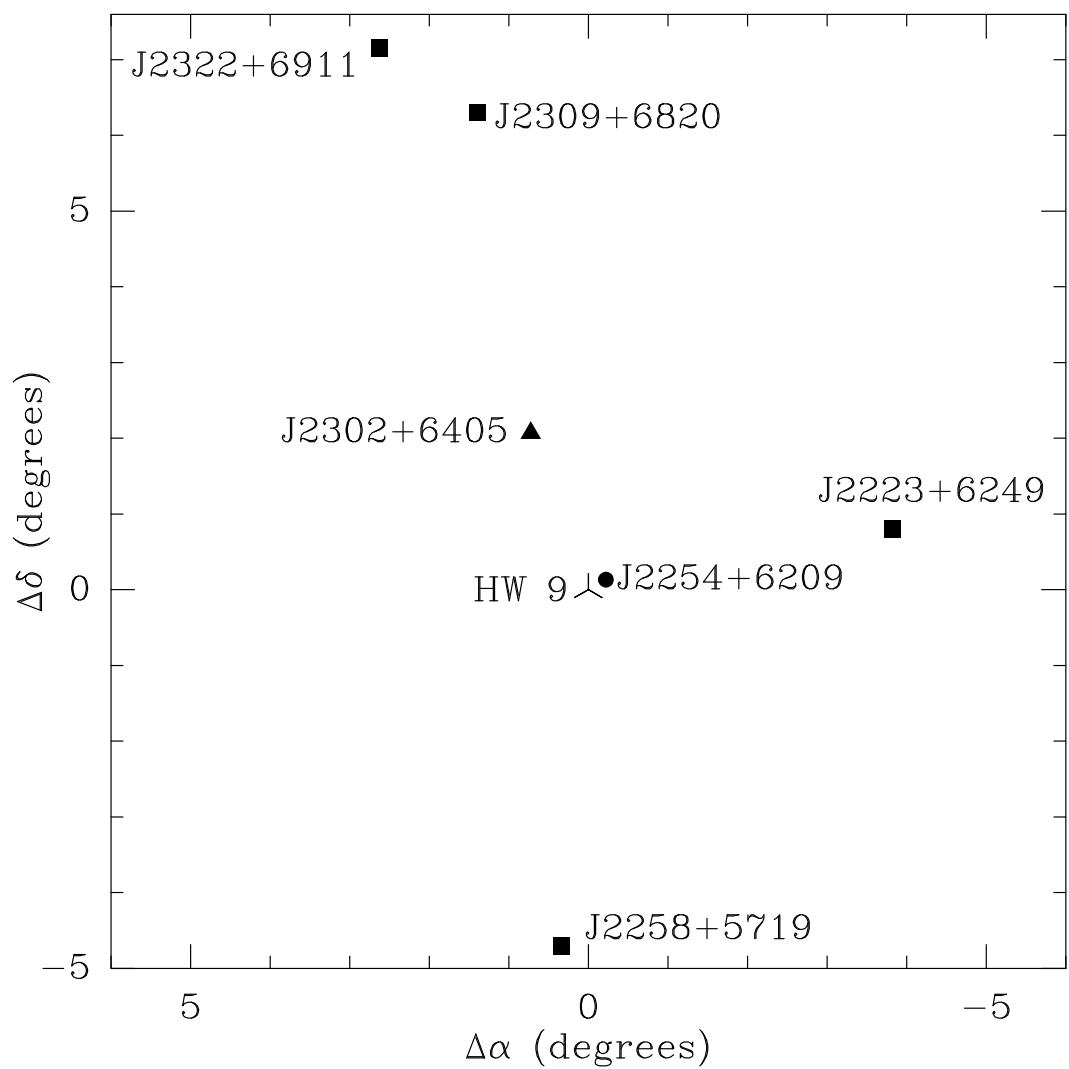}
\end{center}
\caption{Positions of the main and secondary calibrators relative to the target
position, they are plotted as a solid triangle and solid squares, respectively.
The faint quasar J2254+6209 was used to check on the overall quality
of the astrometry, and is shown as a solid circle.}
\label{fig:calibrators}
\end{figure}

In total, ten continuum observations collected at a wavelength of 3.6 {\rm cm} ($\nu$ = 8.42 {\rm GHz})
will be reported here. The first one (obtained in February 2007) was designed as a detection 
experiment. Following the successful detection of the source in that first observation, we initiated 
a series of nine observations starting in October 2007. The separation between 
successive observations in this subsequent data set was about three months, so the last observation 
occurred in November 2009. Our main phase calibrator for all epochs was J2302+6405, 
located at an angular distance of 2.19 degrees from the target. 
To improve the quality of phase calibration, we also observed secondary phase
calibrators. For the first epoch we used the quasars J2258+5719, J2223+6249, and J2322+6911 (at angular
distances of 4.72, 3.90 and 7.63 degrees, respectively, from the target). For the last nine observations,
J2322+6911 was replaced by the somewhat more nearby quasar J2309+6820 at 6.45 degrees from the
target (see Figure \ref{fig:calibrators} for the relative positions of the calibrators). The faint quasar
J2254+6209, located at 0.26 degrees from HW 9 was also observed 
during the last nine observations. It could not be used as a calibrator because it is faint, 
resolved, and variable, but provided a 
very useful check on the overall quality of the astrometry.

Each observation consisted of series of cycles with two minutes spent on source, and one minute 
spent on the main phase calibrator J2302+6405. Roughly every 30 minutes, we observed the 
secondary calibrators, spending one minute on each. In addition, geodetic blocks consisting of
observations of about two dozen calibrators spread over the entire visible sky were collected at 
the beginning, the middle, and the end of each of our multi-epoch observations. With these
overheads, five and three hours were spent on source during the first observation, and each
of the subsequent epochs, respectively. The data were edited and calibrated using the
Astronomical Image Processing System (AIPS; Greisen 2003). The basic data reduction 
followed the standard VLBA procedure for phase-referenced observations, including the 
multi-calibrator schemes and tropospheric and clock corrections. These calibrations were 
described in detail by Loinard et al.\ (2007), Torres et al.\ (2007) and in Dzib et al.\ (2010). 
After their calibration, the visibilities were imaged with a pixel size of 50 $\mu$as using 
a weighting scheme intermediate between natural and uniform (ROBUST = 0 in AIPS). The 
r.m.s.\ noise levels in the final images were 0.08 -- 0.12 mJy beam$^{-1}$. From these images,
the source position was determined using a two-dimensional fitting procedure (task JMFIT in 
AIPS).

\section{Results}

HW 9 was detected in four of the ten observed epochs. These detections imply brightness
temperatures reaching 10$^8$ K, and clearly demonstrate that the radio emission is of 
non-thermal (presumably gyrosynchrotron) origin. The source is very variable, reaching a
maximum flux of about 2.8 mJy in the first and eighth epochs (Figure 3) while remaining 
undetectable at levels below $\sim$0.22 mJy in several observations. This corresponds
to a maximum-to-minimum flux ratio in excess of at least 14. The combination of the high brightness
and variability of the radio emission indicates that HW 9 is a flaring star with coronal emission, 
as previously suggested (e.g., Garay et al.\ 1996). 

The positions of HW 9 measured from our VLBA observations were modeled as a combination 
of a trigonometric parallax ($\pi$) and proper motion ($\mu$ --assumed to be uniform) following 
Loinard et al.\ (2007). The barycentric coordinates of the earth as well as the Julian date 
appropriate for each observation were calculated using the Multi-year Interactive Computer 
Almanac (MICA) distributed as a CD ROM by the US Naval Observatory. The reference epoch 
was taken at JD 2454590.95 $\equiv$ J2008.36, the mean epoch of our our detections. The best 
fit to the data (Figure 4) yields the following astrometric elements:

\begin{eqnarray*}
\alpha_{J2008.36} & = & 22^{{\rm h}}56^{{\rm m}}18\rlap.{^{\rm s}}64308\pm 0.00001\\
\delta_{J2008.36} & = &\ 62^{\circ}1^{'}47\rlap.{''}83902 \pm 0.00005\\
\mu_\alpha \cos{\delta}&=&-0.76 \pm 0.11\ {\rm mas\ yr}^{-1}\\
\mu_\delta&=&-1.85 \pm 0.04\ {\rm mas\ yr}^{-1}\\
\pi&=&1.43 \pm0.07\ {\rm mas}
\end{eqnarray*}

The post-fit r.m.s.\ is 0.04 mas in declination, indicating that no systematic errors remain along 
that axis. In right ascension, however, a systematic contribution of 0.15 mas has to be added 
quadratically to the errors given by JMFIT to yield a reduced $\chi^2$  of 1. These systematic 
errors might be related to the east-west structure present in the images of the main phase 
calibrator, and have been included in all the uncertainties quoted in the present paper.

\begin{figure}[!t]
\begin{center}
\includegraphics[width=0.8\linewidth,angle=-90]{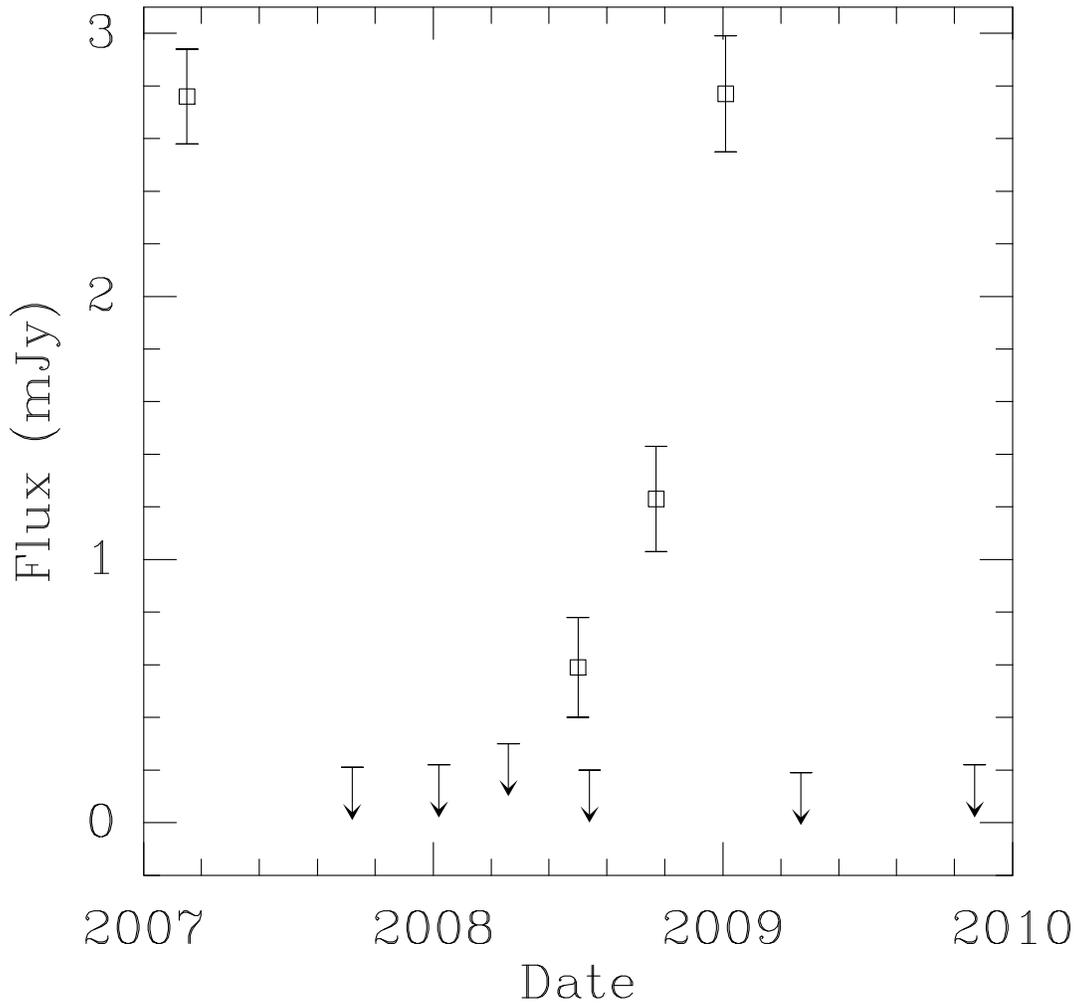}
\end{center}
\label{fig:flux}
\caption{Radio fluxes at 3.6 cm of HW 9 during the ten observed epochs. The detections
were obtained at epochs 2007.20, 2008.50, 2009.77 and 2009.01. The upper limits 
correspond to 3$\sigma$.}
\end{figure}

\section{Discussion}

\subsection{Distance and proper motion of Cepheus A}

The parallax found here for HW 9 is in excellent agreement with the value found by Moscadelli et 
al.\ (2009) for a methanol maser associated with the source HW 2 in Cepheus A ($\pi$ = 1.43 $\pm$ 
0.08 mas). This concordance between two independent results confirms the reliability of VLBA (and, 
more generally, of very long baseline interferometry) measurements of trigonometric parallaxes. By 
combining the two results, we constrain the parallax of the Cepheus A region to be 1.43 $\pm$ 0.06, 
corresponding to a distance $d$ = 700$^{+31}_{-28}$ pc.

The astrometry carried out by Moscadelli et al.\ (2009) yielded accurate proper motions for a methanol 
maser associated with the massive young stellar object HW 2. The velocity of such masers are usually 
believed to agree with that of their associated young stars to better than 3 km s$^{-1}$  (Moscadelli et 
al.\ 2002, 2009). The best proper motion obtained by Moscadelli et al.\ (2009) for the methanol maser 
in the HW 2 region was $\mu_{\alpha}\cos{\delta}=0.5\pm1.1$ mas yr$^{-1}$ and $\mu_{\delta}=-3.7\pm0.2$ 
mas yr$^{-1}$.  
By comparing these figures with our own results (Section 3), it can be seen that the proper motions of 
HW 2 and HW 9 are consistent within $1\sigma$ in right ascension, but differ by more than $3\sigma$ 
in declination (see Figure \ref{fig:pm}). The difference ($\Delta \mu_{\delta}$)
corresponds to a velocity difference $\Delta v$ = 6.2 $\pm$ 0.7 km s$^{-1}$, which likely reflects the
combination of (i) a $\sim$ 3 km s$^{-1}$ difference between the velocity of HW 2 and that of its
associated maser, and (ii) a few km s$^{-1}$ difference between the space velocities of HW 2 and 
HW 9 due to the expected velocity dispersion within the Cepheus A region.

\begin{figure}[!ht]
\begin{center}
\includegraphics[width=0.7\linewidth,angle=-90]{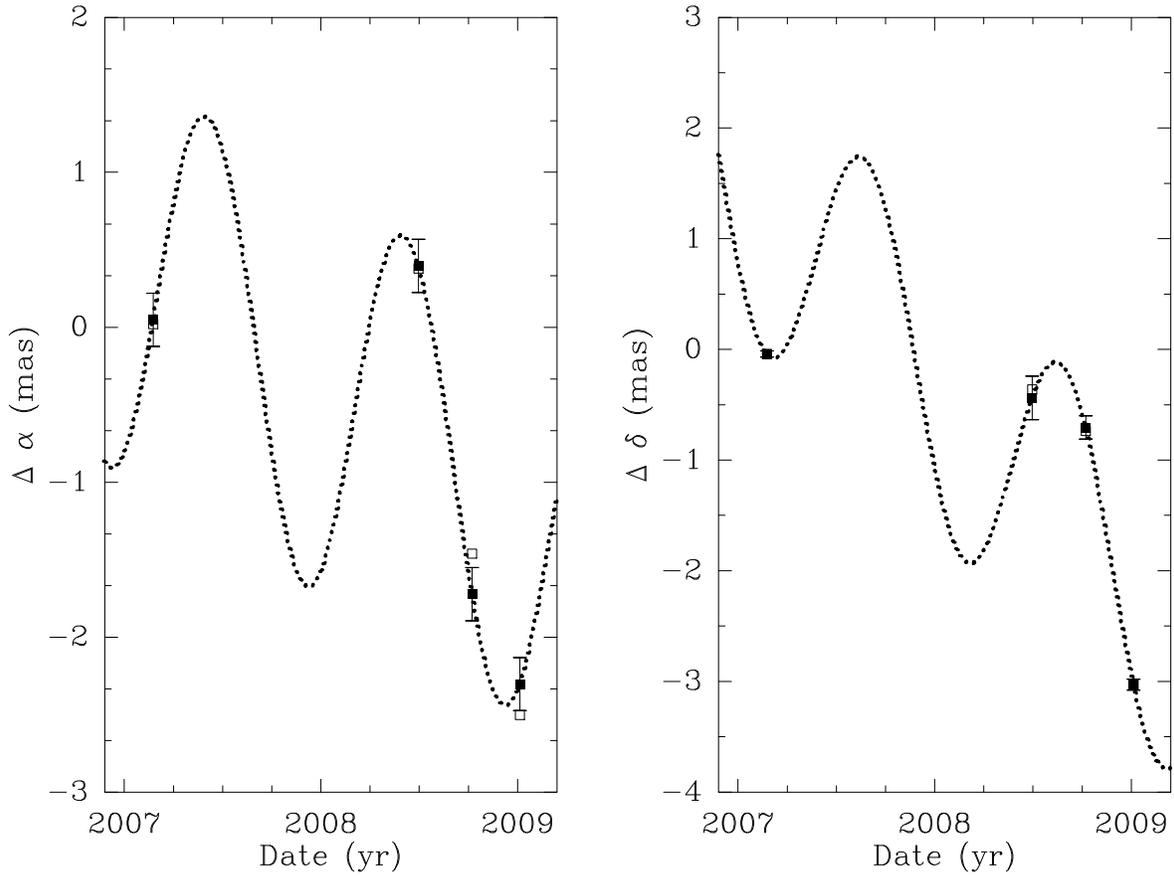}
\end{center}
\caption{Observed positions (open squares) in right ascension (left panel) and declination (right panel)
for the four epochs when HW 9 was detected. The dotted curves show the best fit with a combination of
trigonometric parallax and uniform proper motion, and the filled squares show the expected position of
HW 9 at the four detected epochs according to the best fit.}
\label{fig:fit}
\end{figure}

\subsection{Depth of the Cepheus--Cassiopeia complex}

Using VERA (VLBI Exploration of Radio Astrometry) observations of water masers, Hirota et al.\ (2008)  
estimated the distance ($d$ = 764 $\pm$ 27 pc) to the young massive stellar object IRAS$22198+6336$. 
In projection, this source is located at 4.4 degrees (about 55 pc) from Cepheus A and both regions are 
part of the Cepheus--Cassiopeia molecular cloud complex. The difference in distance appears to be 
similar to the separation on the plane of the sky, suggesting that  the Cepheus--Cassiopeia is about as
deep as it is wide. We note that similar depths have been found for other molecular cloud complexes 
(i.e. Strai{\v z}ys et al.\ 2003)

\subsection{On the nature of HW 9}

As mentioned in Section 1, the exact nature of HW 9 remains unclear. The present detection
with the VLBA clearly demonstrates that the radio emission is of non-thermal origin, and traces 
a flaring corona. This is in agreement with previous analyses of VLA observations 
(e.g.\ Garay et al.\ 1996), and with the interpretation proposed by Pravdo et al.\ (2009) of the 
X-ray emission toward HW 9. High-mass stars are not expected to power active coronas 
because they are fully radiative. As a consequence, the dynamo mechanism cannot operate 
in their interior and they do not generate the strong magnetic fields required to maintain a
corona\footnote{The star S1 in Ophiuchus is a fairly massive B4 star which does generate 
non-thermal radio emission easily detectable with the VLBA. That radio emission, however, 
appears to show only very moderate variability (unlike that of HW 9) and results from a different
type of activity  (Loinard et al.\ 2008; Andr\'e et al.\ 1988, 1991).}. Our observations, therefore, 
indicate that HW 9 is either a low mass (T Tauri) or an intermediate mass (Herbig Ae/Be) young 
star.

For T Tauri stars, the X-ray and radio emissions appear to be well correlated (e.g.\ Benz \&
G\"udel 1994). The typical X-ray to radio luminosity ratio for these objects is $L_X/L_R$ 
$\sim$ 10$^{15.5}$ Hz, with a dispersion around the relation of about one dex (G\"udel 2004; 
Benz \& G\"udel 1994). For Herbig Ae/Be stars, on the other hand, Hamidouche et al.\ (2008) 
found a relation $L_{X}/L_{R}\sim10^{11}-10^{12}$ Hz, but with significant dispersion. In 
particular, values of  $L_{X}/L_{R}$ as high as $10^{13}$ Hz were found. Using radio 
observations with lower angular resolution than those presented here, Pravdo et al.\ (2009) 
found $L_{X}/L_{R}=(0.2-1.4)\times 10^{14}$ Hz for HW 9. Combining the data presented here 
with the X-ray observations of Pravdo et al.\ (2009), we find a similar $L_{X}/L_{R}$ ratio of 
$(0.1-0.8)\times10^{14}$ Hz. This places HW 9 near the lower-end of the $L_{X}/L_{R}$ 
relation for T Tauri stars, and near the upper-end of the relation for Herbig Ae/Be stars.
Our observations, therefore, do not strongly constrain the mass of HW 9 beyond the fact that
it is not a massive object.

The Cepheus A region is a site of very recent star-formation; HW 2, in particular, is believed to be
a very young stellar object. The source HW 9 studied here is located only 5$''$ (less than 0.02 pc) 
away, and is very likely to be coeval with HW 2. The very high extinction toward HW 9 mentioned 
in Section 1 further reinforces the idea that it is a highly embedded, very young object. Indeed, 
Pravdo et al.\ (2009) suggested that HW 9 might be a Class 0/I proto-stellar object. Evidence for 
magnetic activity around such young objects has been reported in a limited number of cases 
(see G\"udel 2002, for a discussion). We note, however, that coronal radio emission has never 
been detected from Class~0 sources, and has been conclusively established for only a few 
Class~I objects (Forbrich et al.\ 2009; Dzib et al.\ 2010). The likely reason for this paucity is related 
to the existence around very young objects of partially ionized winds which generate optically 
thick free-free radio emission surrounding the young stellar source. Any non-thermal coronal radio 
emission from the young star itself would be absorbed in the optically thick layers of the winds, and 
would never reach the observer. The amount of free-free emission associated with HW 9, however, 
appears to be very limited. In particular, VLA observations at 6 cm have sometimes failed to detect 
emission from HW 9 at levels of about 0.15 mJy (Hughes \& Wouterloot 1984; Garay et al.\ 1996). 
Thus, if HW 9 is indeed a very young object, it is one with very limited ejection activity, and this
lack of strong winds would help explain the presence of the coronal emission detected here.

\section{Conclusions}

In this paper, we reported on VLBA observations of the young stellar object HW 9 in the
Cepheus A star-forming region. These data have been used to provide an independent 
confirmation that Cepheus A is located at a distance of 700 pc, and to reduce the uncertainty 
on that distance from about 40 pc down to about 30 pc. While it  is clear that HW 9 is young 
and less massive than $\sim$ 6 \Msun, its exact nature remains unclear. In particular, its radio 
properties are intermediate between those of low-mass T Tauri stars and those of intermediate 
mass Herbig Ae/Be objects. 

\acknowledgments
L.L.\ is grateful to the Guggenheim foundation for financial support. S.D., L.L. and L.F.R.\ 
acknowledge the financial support of DGAPA, UNAM and CONACyT, M\'exico, while R.M.T.\ 
acknowledges support by the Deutsche Forschungsgemeinschaft (DFG) through the Emmy 
Noether Research grant VL 61/3-1. The National Radio Astronomy Observatory is a facility of 
the National Science Foundation operated under cooperative agreement by Associated 
Universities, Inc.

\end{document}